\documentclass[reprint,amsmath,amssymb,aps,prb,]{revtex4-1}

\usepackage{graphicx}
\usepackage{dcolumn}
\usepackage{bm}
\usepackage{subfigure}

\begin{document}
	\preprint{APS/123-QED}
	
	\title{Muon spin rotation and relaxation study on Nb$_{1-y}$Fe$_{2+y}$}
	
	\author{J. Willwater$^1$, D. Eppers$^{1,2}$, T. Kimmel$^1$, E. Sadrollahi$^{1,3}$, F. J. Litterst$^1$, F. M. Grosche$^4$, C. Baines$^5$ and S. S\"ullow$^1$}

	\affiliation{$^1$Institut f\"ur Physik der Kondensierten Materie, TU Braunschweig, D-38106 Braunschweig, Germany\\
	$^2$Physikalisch-Technische Bundesanstalt PTB, Bundesallee 100, D-38116 Braunschweig, Germany\\
	$^3$Institut für Festk\"orper- und Materialphysik, Technische Universit\"at Dresden, D-01069 Dresden, Germany\\
	$^4$Cavendish Laboratory, University of Cambridge, Cambridge CB3 0HE, United Kingdom\\
	$^5$Laboratory for Muon Spin Spectroscopy, Paul Scherrer Institut, CH-5232 Villigen PSI, Switzerland}
	
	\date{\today}
	
	\begin{abstract}
	We present a detailed study of the magnetic properties of weakly ferromagnetic/quantum critical Nb$_{1-y}$Fe$_{2+y}$ using muon spin rotation and relaxation ($\mu$SR). By means of an angular dependent study of the muon spin rotation signal in applied magnetic fields on a single crystal in the paramagnetic state we establish the muon stopping site in the crystallographic lattice of NbFe$_2$. With this knowledge we develop models to describe the muon spin rotation and relaxation signals in the weakly ferromagnetic, spin density wave and quantum critical phases of Nb$_{1-y}$Fe$_{2+y}$ and fit the corresponding experimental data. In particular, we quantify the $\mu$SR response for quantum critical behavior in Nb$_{1.0117}$Fe$_{1.9883}$ and extract the influence of residual weak structural disorder. From our analysis, Nb$_{1-y}$Fe$_{2+y}$ emerges to be uniquely suited to study quantum criticality close to weak itinerant ferromagnetic order.
	\end{abstract}
	
	\pacs{Valid PACS appear here}
	\maketitle
	
	\section{\label{sec:level1}Introduction}
	
	The study of metallic materials with a ferromagnetic quantum critical transition has been pursued intensively throughout decades \cite{Stewart2001,Vojta2003,Loehneysen2007,Brando2016}. In the course of these studies, a plethora of novel and exotic behavior and phases have been revealed \cite{Vojta2003,Loehneysen2007,Brando2016}. In this field, experimental observations are complemented and vindicated by theoretical modeling, leading to a very detailed understanding of the fundamental concepts underlying this type of quantum criticality. 
	
	More specifically, for clean band ferromagnets it was argued that there are in particular two possible scenarios approaching a quantum critical point \cite{Brando2016,Belitz2005,Chubukov2004}: In the first scenario the second order ferromagnetic phase transition becomes a first order transition upon approaching quantum criticality, as it was shown for example in UGe$_2$ \cite{Pfleiderer2002} and ZrZn$_2$ \cite{Uhlarz2004}. In the second scenario the nature of the low temperature magnetic phase changes from ferromagnetic into a modulated magnetic order like a long-wavelength spin density wave. Recently proposed examples for this second scenario are CeFePO \cite{Lausberg2012}, YbRh$_2$Si$_2$ \cite{Lausberg2013}, PrPtAl \cite{Abdul-Jabbar2015} and LaCrGe$_3$ \cite{Taufour2016}. 
	
	\begin{figure}[t]
		\includegraphics[scale=0.34]{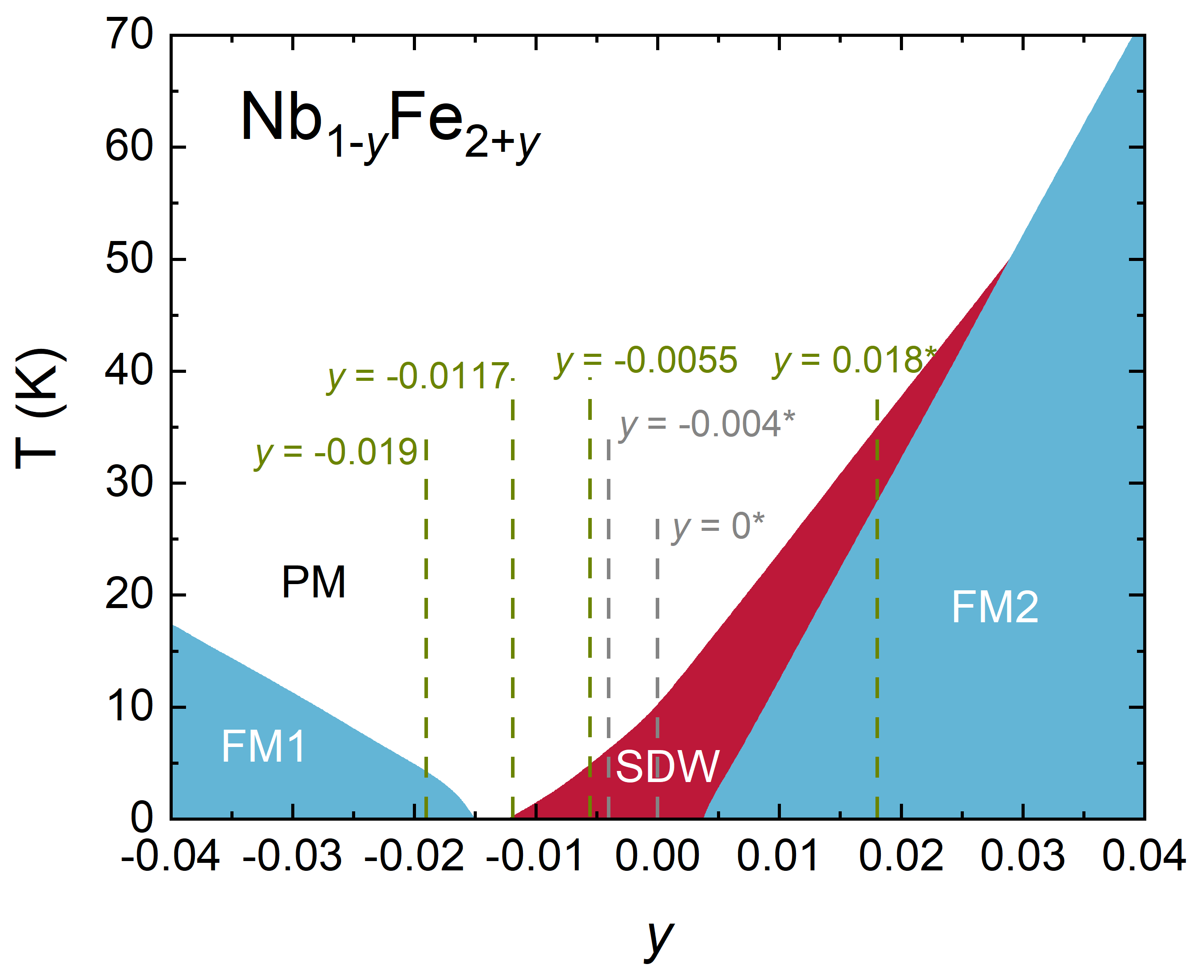}
		\caption{\label{PD} Phase diagram of Nb$_{1-y}$Fe$_{2+y}$ with the ferromagnetic ultralow-moment phases (FM1, FM2) and the spin density phase (SDW) after Niklowitz \textit{et al.} \cite{Niklowitz2019}. In addition, we indicate the samples that are investigated in this paper (green dashed lines) or were investigated previously by means of $\mu$SR \cite{Rauch2015} (marked with *); for details, see text.}
	\end{figure}

	In the context of the latter scenario, a unique example is the itinerant magnet NbFe$_2$ \cite{Shiga1987,Yamada1988,Crook1995,Brando2008,Moroni-Klementowicz2009}. The material crystallizes in the hexagonal MgZn$_2$ (C14) Laves phase \cite{Tompsett2010} with the lattice constants $a=4.8401(2)$\,\AA ~and $c=7.8963(6)$\,\AA . The magnetic phases of Nb$_{1-y}$Fe$_{2+y}$ at low temperatures exhibit a strong sensitivity on the actual sample composition, resulting in a complex magnetic phase diagram \cite{Shiga1987,Yamada1988,Crook1995,Brando2008}. Extensive investigations in recent years using different measurement techniques have established that Nb$_{1-y}$Fe$_{2+y}$ can be pushed into a quantum critical point (QCP) by variation of $y$ \cite{Brando2008,Moroni-Klementowicz2009,Tompsett2010,Rauch2015,Margineda2017,Niklowitz2019,Friedemann2017}, with the phase diagram depicted in Fig. \ref{PD}. From measurements of the magnetic properties on Nb-rich ($y<-0.02$) samples a ferromagnetic ground state was reported \cite{Moroni-Klementowicz2009}. For this ferromagnetic phase FM1, the Curie temperature decreases with increasing Fe concentration until the magnetic order is fully suppressed. Up to $y\sim-0.012$ no long range magnetic order is observed, with signatures of quantum critical behavior occurring for instance in the resistivity and specific heat. Upon further increasing the iron concentration, beginning at lowest temperatures, the samples show signs of a spin-density-wave (SDW) state \cite{Brando2008}. For compositions with $y>0.004$ again a ferromagnetic ground state (FM2) was detected with increasing Curie temperatures for increasing Fe concentration. By now, these findings have been validated by a multitude of experimental techniques including microscopic probes such as muon spin relaxation and elastic neutron scattering. \cite{Rauch2015,Margineda2017,Niklowitz2019}. It was concluded, that the ferromagnetic phases have a magnetic moment below $0.1$\,$\mathrm{\mu_B}$/Fe atom \cite{Moroni-Klementowicz2009,Rauch2015,Niklowitz2019}. The transport and thermodynamic properties have been discussed within Fermi-liquid theory and the concomitant breakdown in the vicinity of suppression of magnetic order \cite{Moroni-Klementowicz2009}. 
	
	Based on these observations, it was proposed that the FM quantum critical point in NbFe$_2$ is preempted by SDW order \cite{Niklowitz2019,Friedemann2017}. Moreover, additional magnetic, electric transport and thermal expansion measurements in connection with a two-order-parameter Landau theory modulation show that the FM transition becomes first order in the presence of the SDW phase \cite{Friedemann2017}. It has been proposed that this leads to tricritical points and ``phase wings'' in magnetic fields \cite{Friedemann2017}, as it was proposed by theory \cite{Belitz2005,Belitz1999}.
	 
	Of special interest are the compositions, where the ordering temperature of the SDW becomes zero and the resistivity and heat capacity show a logarithmic Fermi-liquid breakdown \cite{Brando2008}. Correspondingly, Brando \textit{et al.} described this point in the phase diagram as SDW quantum critical point \cite{Brando2008}. The ``true'' ferromagnetic QCP at higher $y$ is then excluded by the SDW order \cite{Niklowitz2019,Friedemann2017}. In this situation, it calls for further detailed characterization in particular by microscopic experimental techniques of the physical behavior of Nb$_{1-y}$Fe$_{2+y}$ to classify and quantify potential quantum critical behavior.
	
	Therefore, here, we present detailed muon spin relaxation ($\mu$SR) measurements on a polycrystalline sample Nb$_{1.0117}$Fe$_{1.9883}$, for which thermodynamic studies have neither found evidence for a SDW nor a ferromagnetic transition \cite{Duncan2010}. Regarding $\mu$SR studies, it was previously assumed that there are two distinct muon stopping sites in NbFe$_2$ \cite{Rauch2015,Margineda2017}. In order to verify this statement and to allow a full quantitative analysis of the $\mu$SR-spectra, we have determined the muon positions on a Nb$_{0.982}$Fe$_{2.018}$ single crystal. Finally, the magnetic phase assumed to be ferromagnetic in the Nb-rich regime (FM1 in Fig. \ref{PD}) is also investigated by $\mu$SR experiments on polycrystalline Nb$_{1.019}$Fe$_{1.981}$. Then, the muon response is compared to the behavior on the Fe-rich side of the phase diagram close to quantum criticality in polycrystalline Nb$_{1.0055}$Fe$_{1.9945}$. 
		
	\section{\label{sec:level2}Experimental details and data analysis}
	
	The crystals studied in this paper are from the same growth run as samples reported on in the literature, which were well characterized by various bulk experiments \cite{Moroni-Klementowicz2009,Duncan2010a,Friedemann2013}. Four samples with different compositions (see Fig. \ref{PD}) were analyzed using $\mu$SR: Polycrystals Nb$_{0.982}$Fe$_{2.018}$, Nb$_{1.0055}$Fe$_{1.9945}$, Nb$_{1.0117}$Fe$_{1.9883}$ and Nb$_{1.019}$Fe$_{1.981}$ were prepared by radio-frequency induction melting. Subsequently, from the Nb$_{0.982}$Fe$_{2.018}$ polycrystal a single crystal was grown in an UHV-compatible mirror furnace \cite{Neubauer2011}. The samples were characterized by transport measurements and magnetometry and the single crystal additionally by Laue diffractometry \cite{Duncan2010,Friedemann2013}. A transition from the paramagnetic phase to the SDW phase at 36 K and a FM transition at 32 K were found by thermodynamic studies, M\"ossbauer spectroscopy and $\mu$SR for the Fe-rich Nb$_{0.982}$Fe$_{2.018}$, while no transition was detected for Nb$_{1.0117}$Fe$_{1.9883}$ down to low temperatures ($19$\,mK) \cite{Rauch2015,Duncan2010}. Samples with similar compositions as Nb$_{1.0055}$Fe$_{1.9945}$ showed a magnetic transition in the susceptibility at temperatures around $5$\,K. This was explained by the appearance of the SDW phase \cite{Moroni-Klementowicz2009}. A magnetic transition was also found in thermodynamic properties at $4$\,K for samples with similar compositions as Nb$_{1.019}$Fe$_{1.981}$ and argued to have a ferromagnetic signature \cite{Moroni-Klementowicz2009}.
	
	The muon sites can be determined by studying the depolarization rate in the paramagnetic phase as it was demonstrated for UPd$_2$Al$_3$ \cite{Amato1997}. For this, $\mu$SR experiments in the paramagnetic state, \textit{i.e.}, at $50$\,K, under variation of the crystal direction relative to an applied magnetic field of $0.1$\,T were carried out using the GPS facility of the Swiss Muon Source at the Paul Scherrer Institute, Villigen (PSI). In our case, the oriented sample was rotated around the $c$ axis and we measured the depolarization rate in the $ab$ plane in the angular range between $\theta=120^\circ$ to $\theta=255^\circ$ in $15^\circ$ steps, with $\theta$ the angle between the external magnetic field and the $a$ axis. Additional $\mu$SR experiments were performed at the GPS in zero field (ZF) and with longitudinal field (LF) on the single crystalline sample Nb$_{0.982}$Fe$_{2.018}$ up to $40$\,K, on the polycrystalline sample Nb$_{1.0117}$Fe$_{1.9883}$ up to $16$\,K, and on Nb$_{1.0055}$Fe$_{1.9945}$ and Nb$_{1.019}$Fe$_{1.981}$ up to 12\,K. These measurements are complemented by additional $\mu$SR experiments on Nb$_{1.0117}$Fe$_{1.9883}$ down to lowest temperatures of $19$\,mK using the LTF facility at the PSI.
	
	As is visible from the phase diagram, the measurements cover all the different magnetic phases. To analyze our $\mu$SR spectra, an appropriate fit strategy must be designed. In the following, the formulas describing the muon response in the different phases are listed and the physical meaning is briefly explained. We will start with a measurement in the paramagnetic phase and an applied magnetic field. Then, the muon polarization data can be parameterized with the function
	\begin{align}
		\label{fit1}
		P~=~A \cos \biggl(2 \pi \nu  t+\frac{\pi \phi}{180^\circ} \biggr)e^{-\frac{1}{2}(\sigma t)^2}.
	\end{align}
	The oscillation of the muon polarization reflects the muon precession in the external magnetic field $B_{\mathrm{ext}}$. This is accounted for in our fit by the cosine function with the frequency $\nu$ and the phase $\phi$. The internal field fluctuation triggers a Gaussian damping of the signal with the depolarization rate $\sigma$. 
	
	A central role in our study is played by the ZF measurements. It is common knowledge that the $\mu$SR time spectra taken in a ferromagnetic phase can be described with
	\begin{equation}
		\label{ferro}
		F_{\mathrm{FM}}~=~ \alpha \cos(2 \pi \nu t) e^{-\lambda_{T}t} + (1-\alpha) e^{-\lambda_{L}t}.
	\end{equation}
	Instead, for a SDW phase the cosine is replaced by the zeroth-order Bessel function $j_0$ \cite{Le1993,Le1997}:
	\begin{equation}
		\label{SDW}
		F_{\mathrm{SDW}}~=~\alpha j_0(2 \pi \nu t) e^{-\lambda_{T}t} + (1-\alpha) e^{-\lambda_{L}t}.
	\end{equation}
	In both formulas $\alpha$ is the fraction of the transverse field component, $\lambda_{T}$/$\lambda_{L}$ the transverse/longitudinal damping rate and $\nu$ the frequency. 
	
	The measurements on the quantum critical sample, for which no long-range magnetic order has been observed, are of particular interest. Several effects have to be considered, which can lead to a depolarization of the muons. The common Gaussian Kubo-Toyabe (GKT) function
	\begin{equation}
		\label{GKT}
		F_{\mathrm{GKT}}=\frac{1}{3}+\frac{2}{3} [1-(\sigma t)^2] e^{-\frac{1}{2} (\sigma t)^2 }
	\end{equation}
	represents a mean field distribution due to static nuclear dipoles with the damping rate $\sigma$. Another possible relaxation is caused by field distributions of distant dilute dipole moments of electrons. It can be described by the exponential Kubo-Toyabe function \cite{Amato1997MP}
	\begin{equation}
		\label{LKT}
		F_{\mathrm{LKT}}=\frac{1}{3}+\frac{2}{3} (1-a t) e^{-a t },
	\end{equation}
	with the width of the electronically caused field distribution $a$.
	
	\section{\label{sec:level3}Results of the $\mathbf{\mu}$SR experiments}
	
	\subsection{Muon site in Nb$_{0.982}$Fe$_{2.018}$}
	
	In a first step we determine the muon site in Nb$_{0.982}$Fe$_{2.018}$ from experiments in the paramagnetic phase with varying angles of the external magnetic field relative to the crystal structure.
	
\begin{figure}[t]
		\includegraphics[scale=0.27]{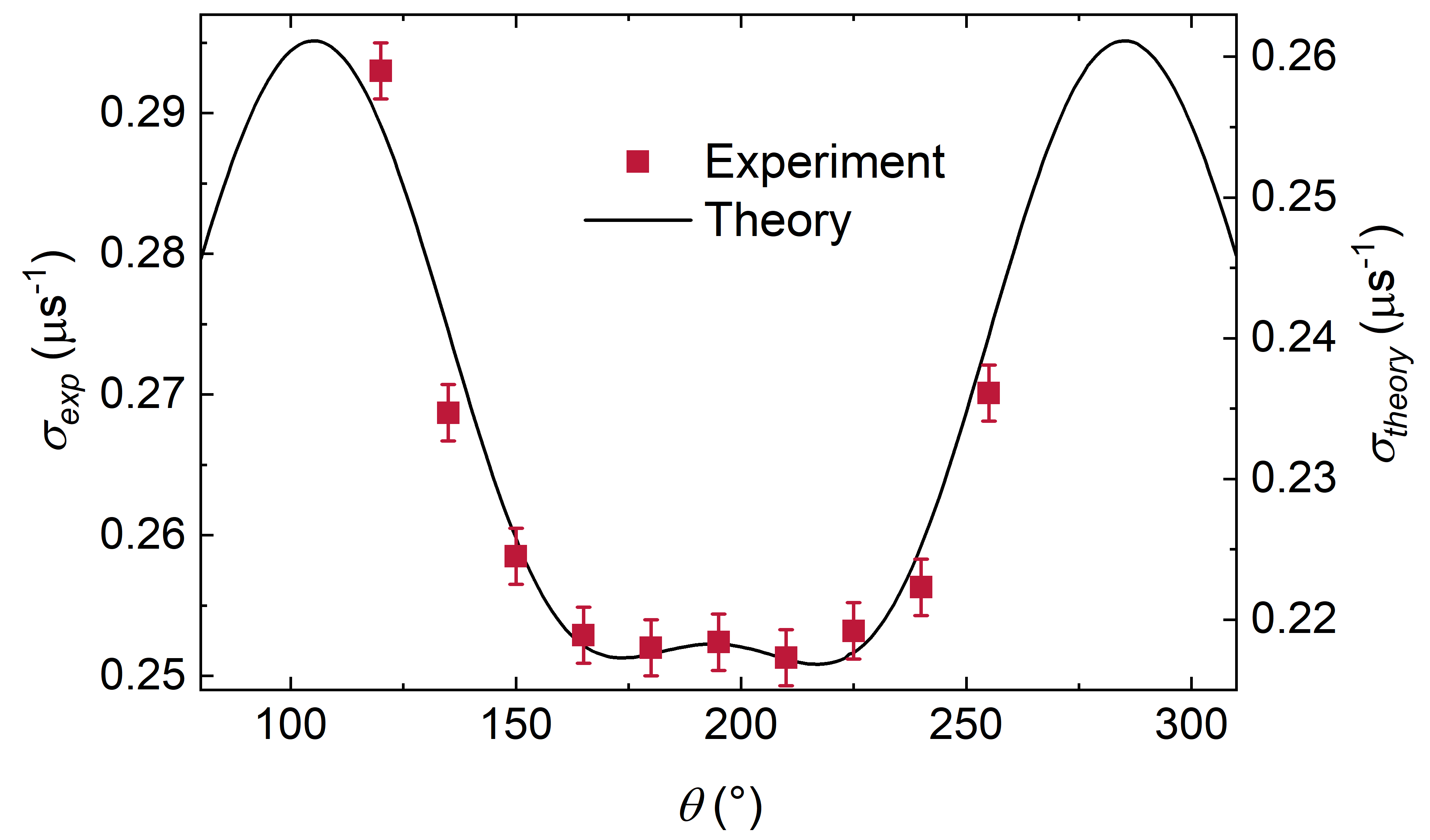}
		\caption{\label{sites} Depolarization rate as function of rotation angle $\theta$ obtained from the experimental data (squares) and simulated (solid line); for details, see text.}
	\end{figure}

	For this the depolarization rate $\sigma$ obtained from the fit with Eq. (\ref{fit1}) is depicted as a function of the rotation angle $\theta$ in Fig. \ref{sites}. The depolarization rate drops with increasing angle up to $\theta=165^\circ$, forms a local maximum at $195^\circ$ and subsequently increases again. In order to associate this behavior to a theoretical model, we describe the internal magnetic field distribution as a sum of dipolar fields over the next four unit cells produced by nuclear moments. The Van-Vleck formula \cite{Vleck1948,Hartmann1977}
	\begin{align}
		\label{Vleck}
		\sigma^2~=~\frac{I_N(I_N+1)}{3} \biggl(\frac{\mu_0}{4\pi} \biggr)^2(\hbar \gamma_\mu \gamma_N)^2\sum_i\frac{(3\cos(\psi_i)^2-1)^2}{r_i^6}
	\end{align}
	is used for the determination of the position. Here, $I_N$ is the nuclear spin, $r_i$ the distance between the muon and the nucleus at site $i$ and $\psi_i$ the angle between $\vec{r_i}$ and the external magnetic field. For a given crystal structure, a fit of the data with the free positional parameters $r_i$ (and the dependent parameters $\psi_i$) thus allows -- at least in principle -- a muon site determination. In practice, usually it is necessary to limit the number of possible muon sites, based for instance on symmetry considerations. Moreover, the muon behaves similar to a hydrogen atom, which occupies interstitial sites in the lattice. The general experience with this ``interstitial-type'' behavior represents another guiding line for identifying possible muon sites.
	
	For NbFe$_2$ a starting point for the simulations are the studies of Gesari \textit{et al.} \cite{Gesari2010}: These authors investigated the position of hydrogen atoms absorbed in compounds that crystallize in the AB$_2$ C14 Laves Phase structure (where A\,=\,Zr, Ti; B\,=\,Ni, Mn, Cr, V). Based on density functional theory calculations it has been demonstrated that the hydrogen atoms prefer tetrahedral positions and the probability of occupation decreases as the hole size decreases. Although NbFe$_2$ was not explicitly investigated in Ref. \cite{Gesari2010}, and in spite of the difference in mass of hydrogen atoms and muons, this report is used to narrow the search area for possible muon sites in the unit cell of NbFe$_2$.
			
	In our simulations, we vary the muon site (and with this $r_i$ and $\psi_i$ in Eq. (\ref{Vleck})) and compare the resulting calculated depolarization rate $\sigma_{theory}$ with the experimentally obtained $\sigma_{exp}$. The optimum solution of our simulation is plotted in Fig. \ref{sites} and was achieved using a muon site with coordinates in the unit cell of $r_{\mu1}\nobreak=\nobreak(0.558(7), 0.278(7), 0.189(5))$. This way, the calculated angular dependence of the depolarization rate shows a good agreement with the experimental data; especially the local extremal points in the angular range between $150^\circ$ to $250^\circ$ are well reproduced. It should be mentioned that there is a constant offset in absolute values between $\sigma_{theory}$ and $\sigma_{exp}$ of $0.034~\mathrm{\mu s^{-1}}$. This difference in absolute values can be explained with the simplifications of the simulation: Only nuclear moments were taken into account and the dipole approximation was used.
	
\begin{figure}[t]
		\includegraphics[scale=0.3]{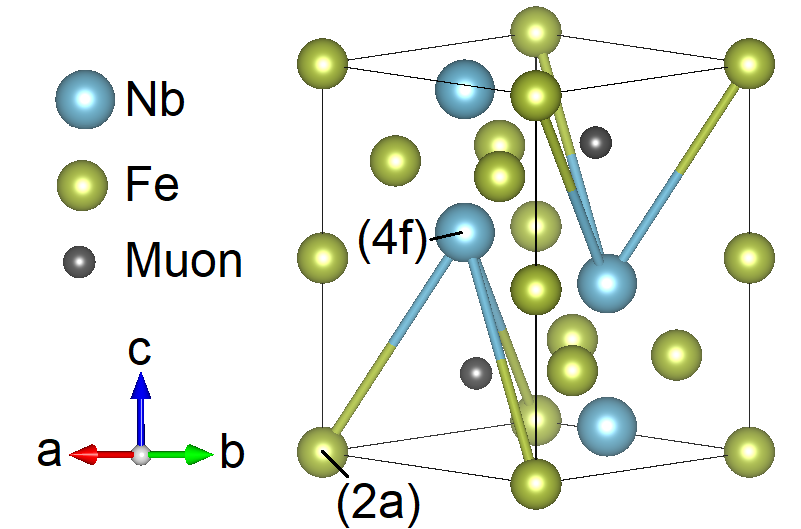}
		\caption{\label{structure} Structure of NbFe$_2$ with $a=4.8401(2)$\,\AA ~and $c=7.8963(6)$\,\AA ~and the calculated muon site; for details see text.}
	\end{figure}

	With the determination of the first muon site, from symmetry considerations, the equivalent muon site is determined to $r_{\mu2}\nobreak=\nobreak(0.441(3), 0.721(3), 0.810(5))$. The equivalent muon sites $r_{\mu1}$ and $r_{\mu2}$ are close to the center of the tetrahedra, which are formed by one Nb(4f) and three Fe(2a) atoms indicated in Fig. \ref{structure}. Notably, there is a slight shift of this position compared to the highly symmetrical places in the tetrahedra (2/3, 1/3, 1/4). 

	We note that ferromagnetic order breaks the inversion symmetry of the crystal structure, and thus lifts the equivalence of the two muon stopping sites. Therefore, in a ferromagnetic phase it will be necessary to fit the muon spectra assuming two distinct, but similar muon stopping sites. Conversely, in a paramagnetic phase at both muon sites there will be essentially the same field and field distribution, and the assumption of a single muon stopping will be sufficient to fit the experimental data (the site occupation in the SDW phase will be discussed in the following subsection). This conclusion is in full accordance with the fit procedures used in previous $\mu$SR studies on Nb$_{1-y}$Fe$_{2+y}$ \cite{Rauch2015,Margineda2017}.
	
	\subsection{FM and SDW phase in Fe-rich Nb$_{0.982}$Fe$_{2.018}$}
	
\begin{figure}[t]
		\includegraphics[scale=0.148]{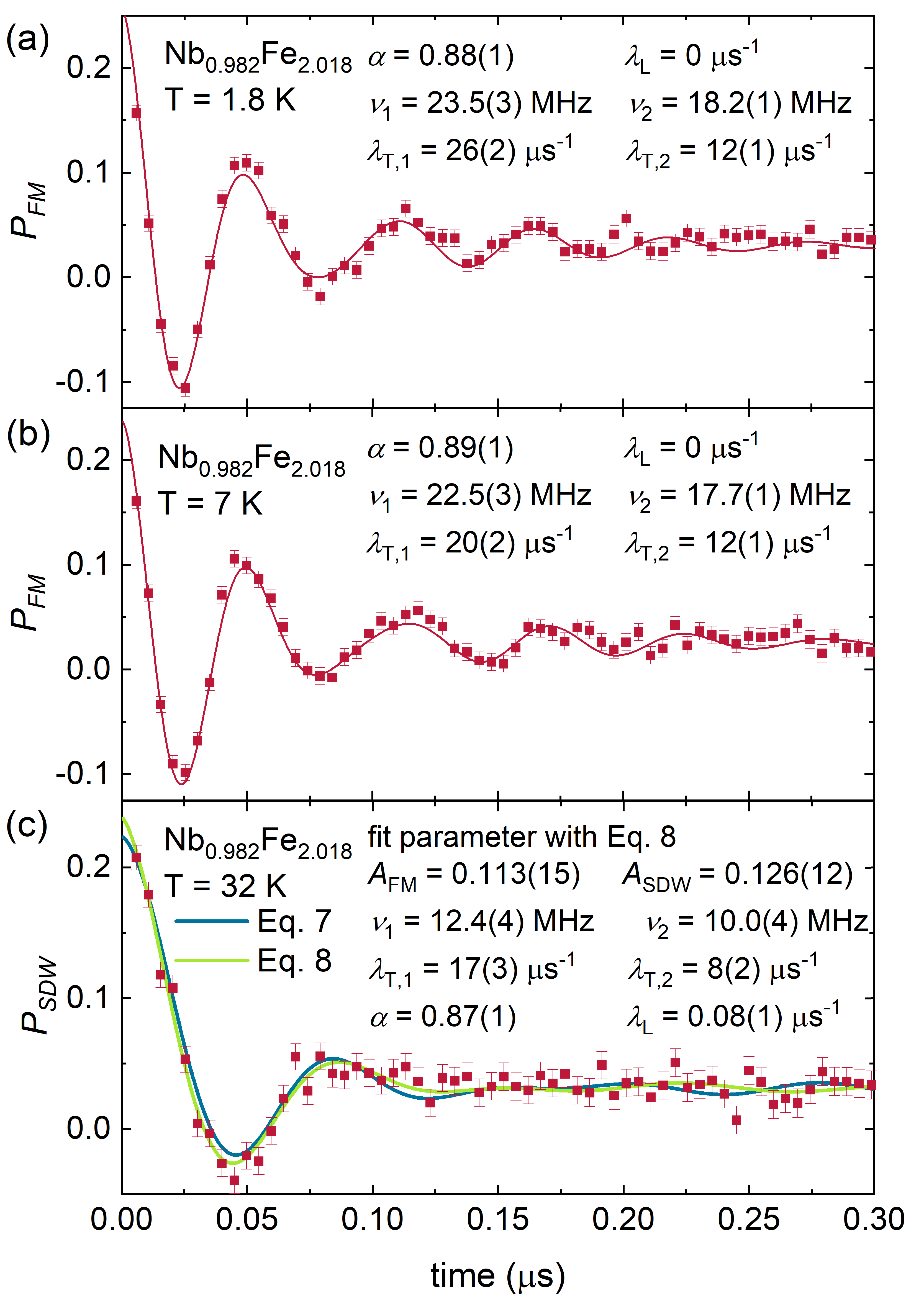}
		\caption{\label{FM_SDW} Zero-field $\mu$SR time spectra of Nb$_{0.982}$Fe$_{2.018}$ for short times in (a) the ferromagnetic regime at $1.8$\,K and (b) $7$\,K and (c) the SDW phase at $32$\,K. The solid lines in (a) and (b) indicate the results of a fit with equation (\ref{fit2}) and in (c) with equations (\ref{fit2}) and (\ref{fit4}); for details, see text.}
	\end{figure}

	With the muon site now established, and to obtain further information about the spin-density-wave phase, we reevaluate the ZF muon spin relaxation experiments on Fe-rich Nb$_{0.982}$Fe$_{2.018}$ from Ref. \cite{Rauch2015} (including a spectral analysis via Fast Fourier Transformation). In an initial step the ZF measurement at $1.8$, $7$\,K and $32$\,K are analyzed. The zero-field $\mu$SR time spectra in Fig. \ref{FM_SDW}(a) and (b) show an oscillating behaviour typical of a ferromagnet. In contrast, the $\mu$SR spectra in the SDW phase exhibits only one oscillation and is then constant (Fig. \ref{FM_SDW}(c)). To verify the occupation probabilities, the same fit approach as in Ref. \cite{Rauch2015} of two muon sites in a ferromagnetic/SDW environment was chosen:
	\begin{equation}
		\label{fit2}
		P_{\mathrm{FM/SDW}}~=~\sum_{i=1}^{2}A_i F_{\mathrm{FM/SDW},i}.
	\end{equation}
	For the two muon sites $i=1,2$ the same fit equation $F_{\mathrm{FM/SDW}}$ [see Eq. (\ref{ferro}) and (\ref{SDW})] with the asymmetry $A_i$ is used. To limit the number of free parameters, the following simplifications are applied: $A_1=A_2$, $\alpha_1=\alpha_2$ and $\lambda_{L,1}=\lambda_{L,2}$. This is justified since the muon sites in NbFe$_2$ are geometrically equivalent. A fit with the equation (\ref{fit2}) describes the experimental data very well (see Fig. \ref{FM_SDW} (a) and (b)), with the fit parameters included in the figure. The two different frequencies and transverse damping rates $\lambda_T$ once again show the need for two muon sites.
	
	In contrast and as demonstrated in Fig. \ref{FM_SDW}(c), the fit with Eq. (\ref{fit2}) does not do a very good job at reproducing the experimental results. The first minimum of the experimentally determined asymmetry spectrum is more pronounced than is modeled by the fit, while the calculated second minimum at $0.12~\mathrm{\mu s}$ is not clearly visible in the data. Therefore, we have tried various alternative approaches. A significantly better matching between experimental data and fit is achieved with a superposition of equations (\ref{ferro}) and (\ref{SDW}) (see Fig. \ref{FM_SDW}(c)): 
	\begin{equation}
		\label{fit4}
		P_{\mathrm{SDW2}}~=~A_{\mathrm{FM}} F_{\mathrm{FM}}+A_{\mathrm{SDW}} F_{\mathrm{SDW}}
	\end{equation}
	With this function the initial steep drop in the asymmetry signal is well reproduced and the fit matches the data well at larger times. Conceptually, this equation describes a coexistence of a ferromagnetic phase [Eq. (\ref{ferro})] with a SDW phase [Eq. (\ref{SDW})], with the asymmetries $A_{\mathrm{FM}}$ and $A_{\mathrm{SDW}}$ a measure of the volume fraction of the two phases. At $32$\,K we obtain values $A_{FM}=0.113(15)$ and $A_{SDW}=0.126(12)$ from the fit. The volume fraction of the SDW phase is thus slightly higher, but comparable to that of the FM phase. We note that here, similar to the FM phase, we should also use two distinct muon sites in the fit. But this would produce a very strong fit parameter interdependency of the phase volume fractions with other parameters, prohibiting to obtain additional meaningful information from such fits. Therefore, we have used only a single averaged muon site instead of two equivalent sites.


	In summary, our refined analysis of the $\mu$SR data from Ref. \cite{Rauch2015} reveals a spatial coexistence of a ferromagnetic phase with a SDW phase in Nb$_{0.982}$Fe$_{2.018}$ at $32$\,K. In line with the published neutron results \cite{Niklowitz2019}, this can be interpreted as a gradual transition from the FM phase into a SDW phase with long wavelength upon increasing temperature. In a more general context, the successful reanalysis of the $\mu$SR data on Nb$_{0.982}$Fe$_{2.018}$ under consideration of the site occupation of the muons allows a more detailed and quantitatively accurate analysis of the full magnetic phase diagram (see below).
	
	A more detailed comparison of the fit parameters in the ferromagnetic and the SDW phase supports the interpretation a gradual transition: While the damping rates should diverge at a critical phase transition, they are almost constant over the gradual FM/SDW phase transition. The frequencies $\nu_i$ are in good agreement with those reported in Ref. \cite{Rauch2015}. The frequencies depend on the local field and can thus be compared with the bulk magnetization. For Nb$_{0.985}$Fe$_{2.015}$ a magnetic moment of approximately $0.06$\,$\mu_\mathrm{B}$/Fe was measured at $7$\,K \cite{Friedemann2013}. Assuming that the magnetic moment in the SDW phase scales with $\nu_i$ in the same way as in the FM phase, we arrive at a magnetic moment of approximately $0.03$\,$\mu_\mathrm{B}$/Fe in the SDW phase. This is fully consistent with the magnetic moment of $<0.1$\,$\mu_\mathrm{B}$/Fe determined in a neutron diffraction experiment \cite{Niklowitz2019}. 
	
	\subsection{Quantum criticality in Nb$_{1.0117}$Fe$_{1.9883}$}
	
\begin{figure}[t]
		\includegraphics[scale=0.1]{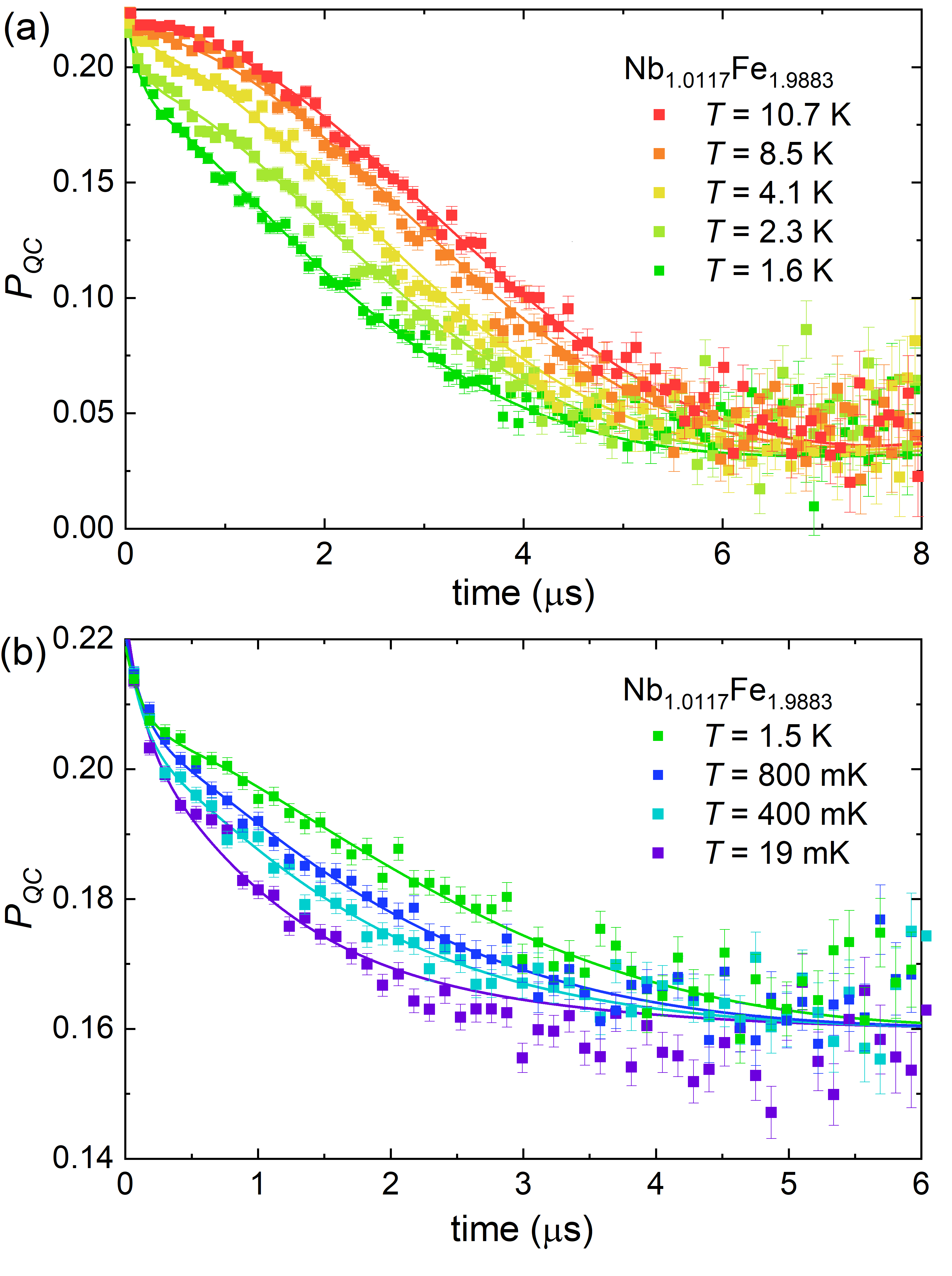}
		\caption{\label{QC} Zero-field $\mathrm{\mu}$SR time spectra of a Nb$_{1.0117}$Fe$_{1.9883}$ polycrystal measured with the (a) GPS instrument in a temperature range from $1.6$\,K to $10.7$\,K and (b) LTF instrument in a temperature range from $19$\,mK to $1.5$\,K. The solid lines denote a fit with equation (\ref{fit_gps}); for details, see text.}
	\end{figure}

	Of particular interest are the $\mu$SR experiments on the polycrystalline Nb$_{1.0117}$Fe$_{1.9883}$ sample which is positioned close to the quantum critical point and for which no long-range magnetic order could be established by other experimental techniques \cite{Duncan2010}. To investigate the nature of the magnetic ground state of this composition, we have carried out ZF and LF $\mu$SR measurements. The ZF measurements performed in a temperature range from $1.6$ to $10.7$\,K (GPS experiment) can be seen in Fig. \ref{QC} (a) and from $19$\,mK to $1.5$\,K (LTF set-up) in Fig. \ref{QC} (b). 
	
	The moderate variation of the asymmetry signal with time clearly demonstrates the absence of long-range magnetic order down to the lowest temperatures. While at higher temperatures $\sim10$\,K $P_{\mathit{QC}}$ reveals an apparent dominant Gaussian damping, it develops an additional more strongly damped contribution at lower temperatures with increasing asymmetry on the expense of the first signal. This type of behavior is clearly visible already in the raw data, e.g., at 1.5\,K [Fig. \ref{QC} (b)]. Moreover, the Gaussian damping character of the first signal transforms continuously into a more exponential one upon lowering temperature. 
	
	Before describing in detail our fit approach, we start with some general observations. In the absence of long-range magnetic order, muons on both stopping sites in the unit cell of Nb$_{1.0117}$Fe$_{1.9883}$ only detect the local magnetic field distributions from magnetic fluctuations (either from the nuclei or electrons) or frozen-in spins or spin-clusters. In result, the experimental data represent an averaging of the magnetic field distribution at the muon stopping sites and the above discussed muon sites can be considered equal. Therefore, unlike the ferromagnetic case, it is sufficient to only assume a single stopping site for the muons.
	
	
	Then, for fitting the experimental data we use a sum of three signal components,
	\begin{equation}
		\label{fit_gps}
		P_{\mathrm{QC}}=A_1 F_{\mathrm{GKT}} e^{-\lambda t} + A_2 F_{\mathrm{GKT}} F_{\mathrm{LKT}} + A_{\mathrm{BG}} .
	\end{equation}	
	The first term of this equation accounts for the different relaxation caused by the dynamics of nuclear and electronic spins. With a static assumed nuclear field distribution and the fast fluctuating electronic ones, the double relaxation can be described by the product of a GKT function for the nuclear contribution and an exponentially damped for the electronic one. The damping of the GKT function was derived from the high temperature spectra and kept constant for all temperatures. A detailed analysis of the temperature behavior of the fluctuating magnetic fields is given below. The second term in Eq. (\ref{fit_gps}) represents the additional damping of the signal picked up upon lowering temperatures and is therefore connected to a different phase volume in the sample. This contribution starts to develop below $11$\,K on the expense of the asymmetry of the first signal and as we show below saturates below $1$\,K. Also here the double relaxation due to the presence of static nuclear fields is taken into account by the same multiplied GKT function as used in the first term of Eq. (\ref{fit_gps}). Its shape is best described by Eq. (\ref{LKT}). The last term $A_{\mathrm{BG}}$  in Eq. (\ref{fit_gps}) stands for an undamped background. For the LTF measurements this is due to a large silver backing of the sample holder, with $A_{\mathrm{BG,LTF}}=0.16$ significantly larger than that for the GPS setup: $A_{\mathrm{BG,GPS}}=0.03$.

	For a quantitative analysis of the relaxation due to fluctuating magnetic fields from the electrons in Nb$_{1.0117}$Fe$_{1.9883}$, we fit all experimental data using Eq. (\ref{fit_gps}), this way deriving in particular the temperature dependence of the asymmetries $A_1$, $A_2$ and the depolarization rate $\lambda$. The parameters $A_1$ and $A_2$ represent the volume fractions of the two different relaxation components, while $\lambda$ quantifies the temperature evolution of the magnetic fluctuations in the sample volume $A_1$. 
	
\begin{figure}[t]
		\includegraphics[scale=0.31]{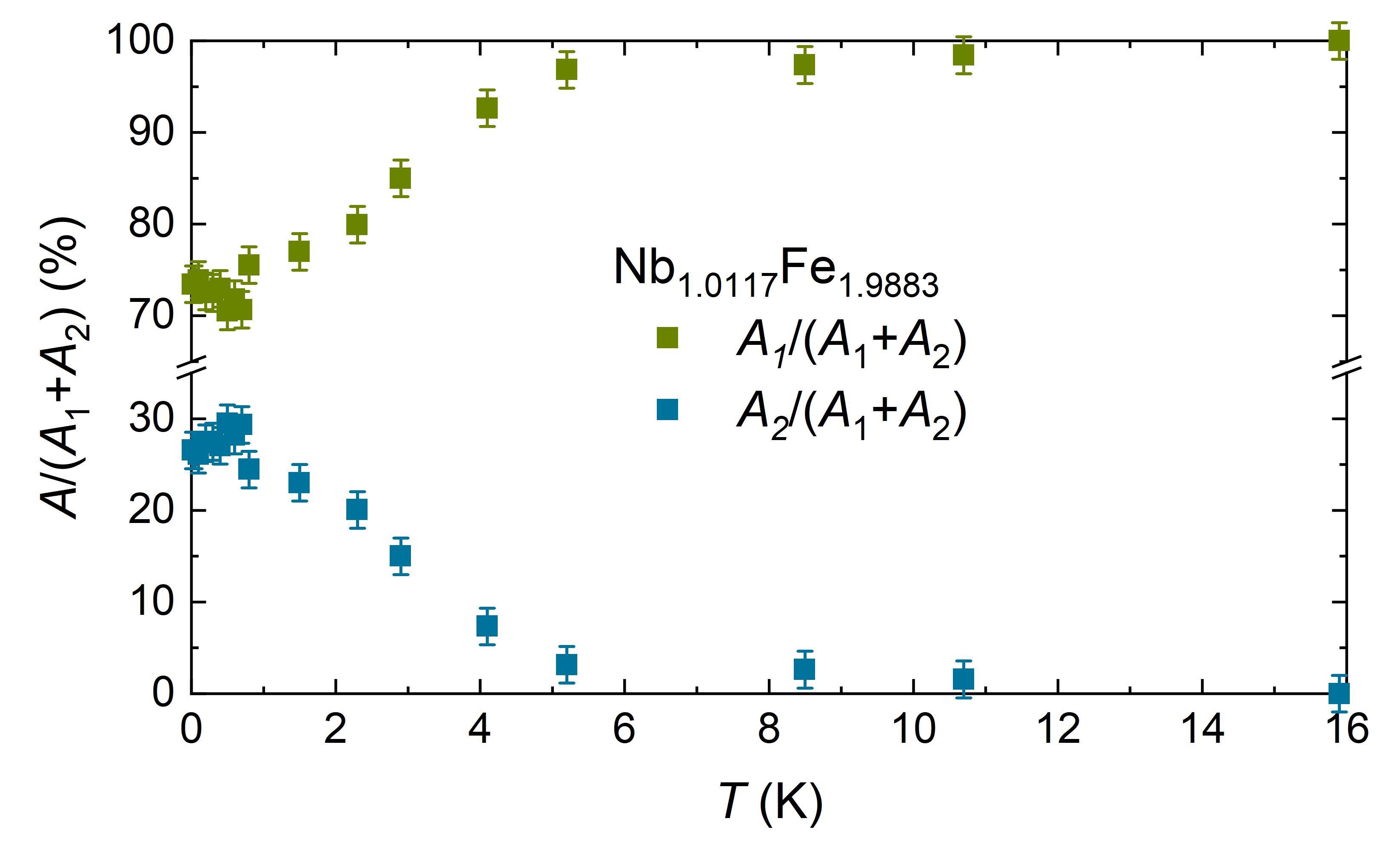}
		\caption{\label{asy} Temperature dependence of the normalized asymmetries $\frac{A_1}{A_1+A_2}$ and $\frac{A_2}{A_1+A_2}$ derived from the $\mu$SR measurements on Nb$_{1.0117}$Fe$_{1.9883}$; for details, see text.}
	\end{figure}

	As a first result of our analysis, we quantify the competition between the two sample volumes associated to the parameters $A_1$ and $A_2$ by plotting the volume fractions $\frac{A_1}{A_1+A_2}$ and $\frac{A_2}{A_1+A_2}$ in Fig. \ref{asy}. At high temperatures the signal is fully described by the first term in Eq. (8) and correspondingly $A_2=0$. Starting at around $11$\,K $A_2$ first rises gradually and below $5$\,K takes off more strongly on the expense of $A_1$. Below $1$\,K, the two volume fractions appear to become constant as function of temperature. In this low temperature range we have thus about $70$\,\% of the signal stemming from the strongly fluctuating paramagnetic $A_1$ range, while in $\sim 30$\,\% of the sample volume the muons encounter a field distribution from dilute dipole moments of electrons.   

\begin{figure}[t!]
		\includegraphics[scale=0.31]{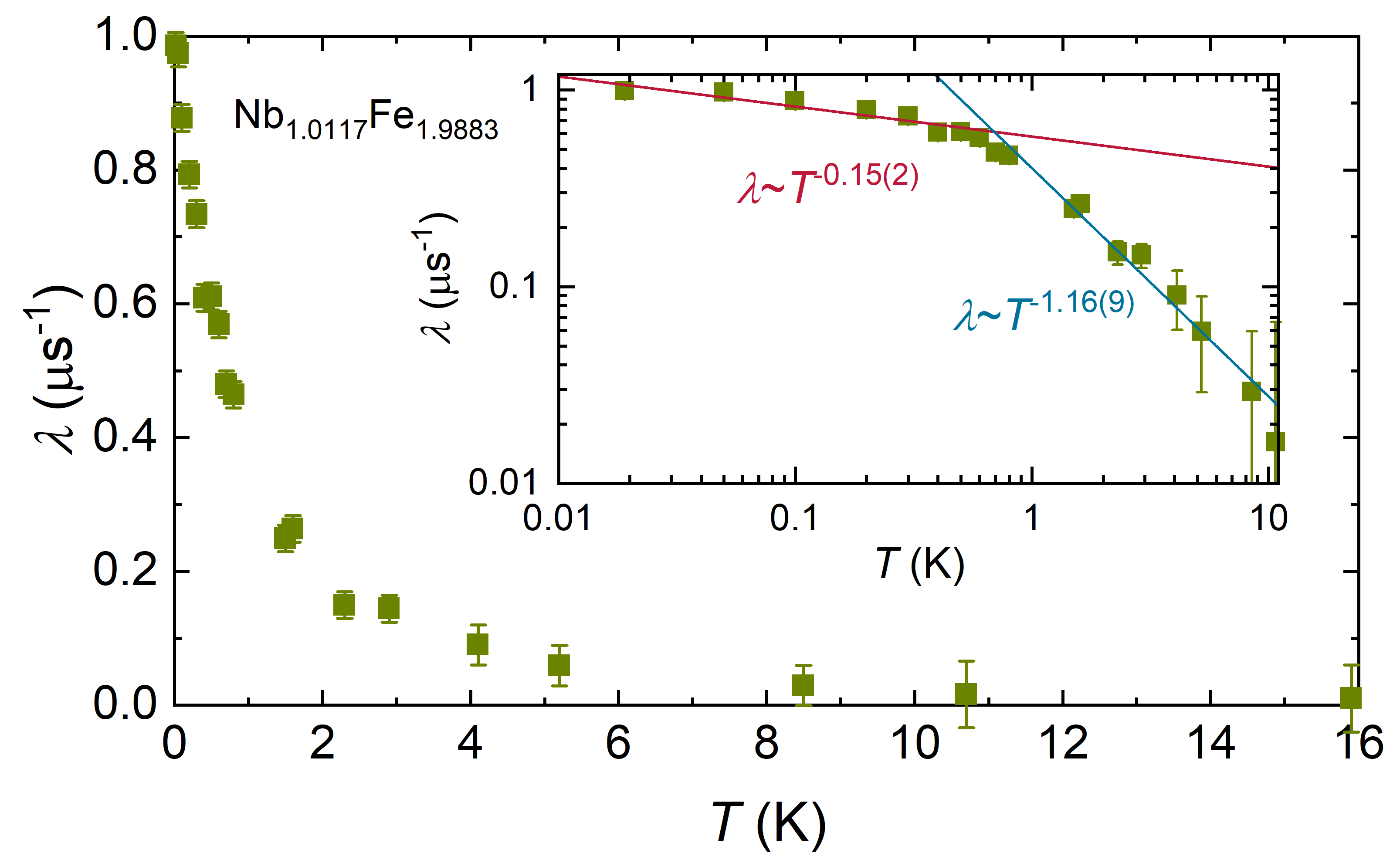}
		\caption{\label{QC_res} Temperature dependence of the depolarization rate $\lambda$, obtained from the muon relaxation measurements on Nb$_{1.0117}$Fe$_{1.9883}$ depicted in Fig. \ref{QC}. The inset shows a double logarithmic plot of $\lambda$ over $T$ with fits (solid lines) according to Eq. (\ref{fit_lambda}); for details, see text.}
	\end{figure}

	Next, the temperature dependence of $\lambda$ is shown in Fig. \ref{QC_res}. At high temperatures the exponential damping is only weak and the depolarization rate $\lambda$ close to zero, since electronic fluctuations are very fast. With decreasing temperature the fluctuations are slowing down, resulting in a stronger damping of the $\mu$SR signal with a more exponential character. It leads to an increase of the depolarization rate $\lambda$. At $\sim 0.7$\,K $\lambda$ crosses over into a different temperature dependence as illustrated in the double logarithmic inset of Fig. \ref{QC_res}. This crossover observation is quantified by power-law fits of the depolarization rate $\lambda$:
	\begin{equation}
		\label{fit_lambda}
		\lambda = c T^{-\alpha}.
	\end{equation}
	At temperatures $> 0.7$\,K a fit of the data yields $c=0.42(3)$\,$\mathrm{\mu s^{-1}}$ and $\alpha=1.16(9)$, while at low temperatures $< 0.7$\,K we find $c=0.58(3)$\,$\mathrm{\mu s^{-1}}$ and $\alpha=0.15(2)$ (inset of Fig. \ref{QC_res} with a double-logarithmic plot of $\lambda (T)$).
	
\begin{figure}[t]
		\includegraphics[scale=0.32]{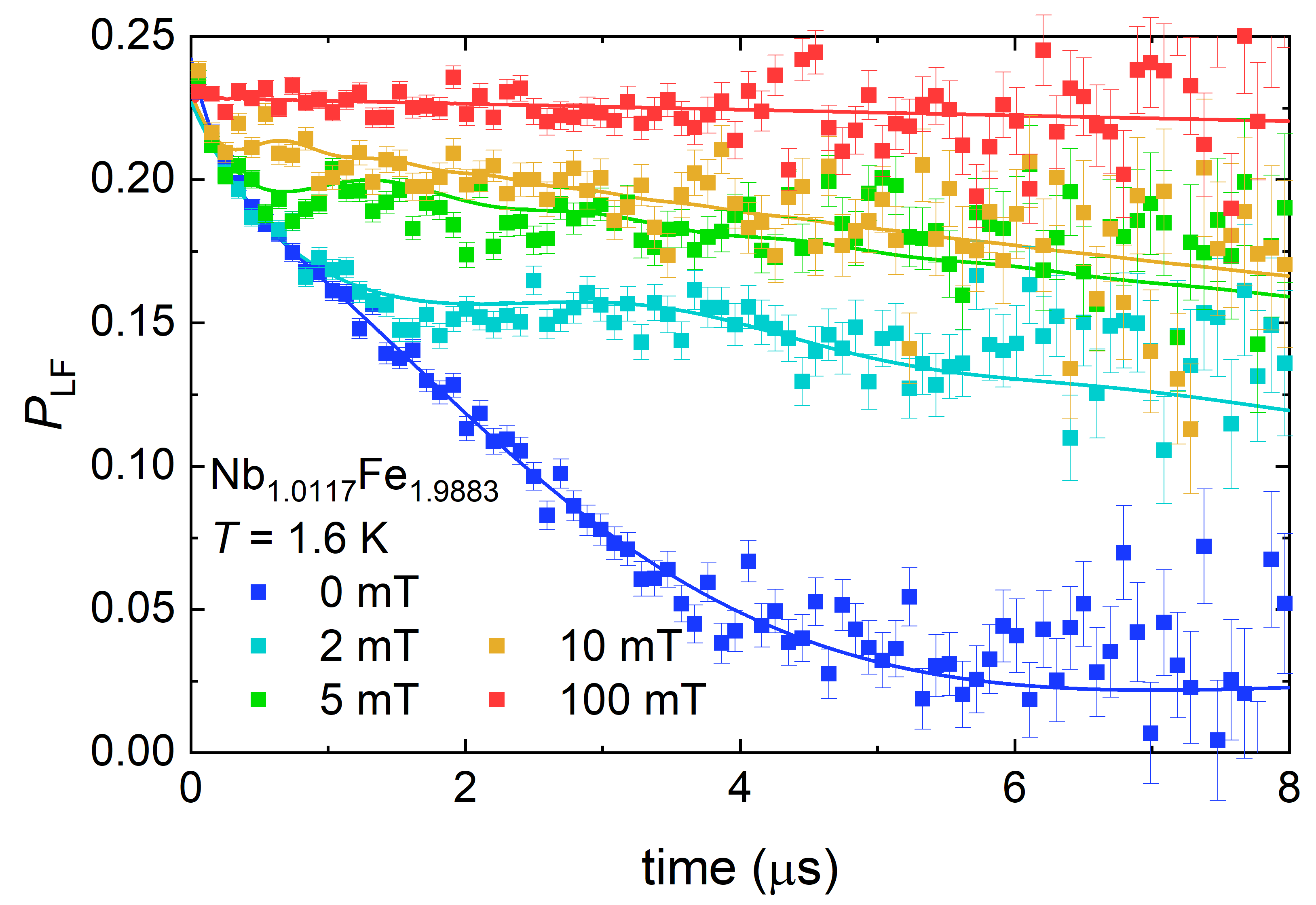}
		\caption{\label{QC_LF} $\mu$SR time spectra of a LF measurement on Nb$_{1.0117}$Fe$_{1.9883}$ for magnetic fields between $0$ and $100$\,mT at $1.6$\,K. The solid lines correspond to fits with Eq. \ref{fit_gps}. Only for the LKT and GKT the static LF functions \cite{Hayano1979,Uemura1985} were used.}
	\end{figure}

		Additional LF measurements shown in Fig. \ref{QC_LF} support the analysis from the ZF measurements: The Gaussian damping by the nuclear moments is readily suppressed by application of longitudinal fields of $2$\,mT proving the static nuclear origin (Larmor precession of the muon spin at this field is already much faster than the nuclear damping rate). This explains the strong difference between the measurements at $0$ and $2$\,mT. The electronic damping is suppressed by about $100$\,mT showing that the evolved field distributions must be small and the fluctuations are already relatively slow (some $10$\,MHz).

	In contrast to the nuclear damping, the LF experiments reveal that the exponential Kubo-Toyabe signal is not suppressed by $2$\,mT since the width of the electronically caused field distribution is large compared to the muonic Larmor precession. This is directly be seen in the rapid decrease of the signal at short times (Fig. \ref{QC_LF}). The situation changes at 10 mT, when the static electronic damping is suppressed and fully overcome by an external field of $100$\,mT.    
	
	Based on the ZF and LF measurements and fits of the quantum-critical sample Nb$_{1.0117}$Fe$_{1.9883}$ a microscopic view of the magnetism at low temperatures emerges. Two different regimes can be distinguished from the temperature dependency of $\lambda$: (i) At high temperatures, the Gaussian like behavior of the $\mathrm{\mu}$SR time spectra indicates a ''conventional'' paramagnetic state. In addition, the increase of the depolarization rate $\lambda$ indicates a slowing down of the fluctuating electronic spins in the quantum critical regime. Down to 0.7\,K $\lambda$ follows the simple proportionality $\lambda \propto T^{-\alpha}$. (ii) In the second region below $0.7$\,K, while still being in a paramagnetic state, a crossover occurs, as the increase of $\lambda$ and the Lorentzian behavior of the $\mathrm{\mu}$SR spectra shows that the fluctuations of the spins slow down even further and the muons are depolarized faster. The temperature dependence of $\lambda$ can be described with the same fit function as above, but with changed parameters. (iii) As temperature is lowered, in a (small) fraction of the sample a different relaxation behavior appears, which has the properties of diluted dipole moments of electrons. 
	
	Our $\mu$SR measurements on the quantum critical sample Nb$_{1.0117}$Fe$_{1.9883}$ are thus fully worked out and parameterized. A detailed interpretation follows in the discussion. 
	
		\subsection{Magnetically ordered phase close to the quantum critical point}
	
	\begin{figure}[t]
		\includegraphics[scale=0.148]{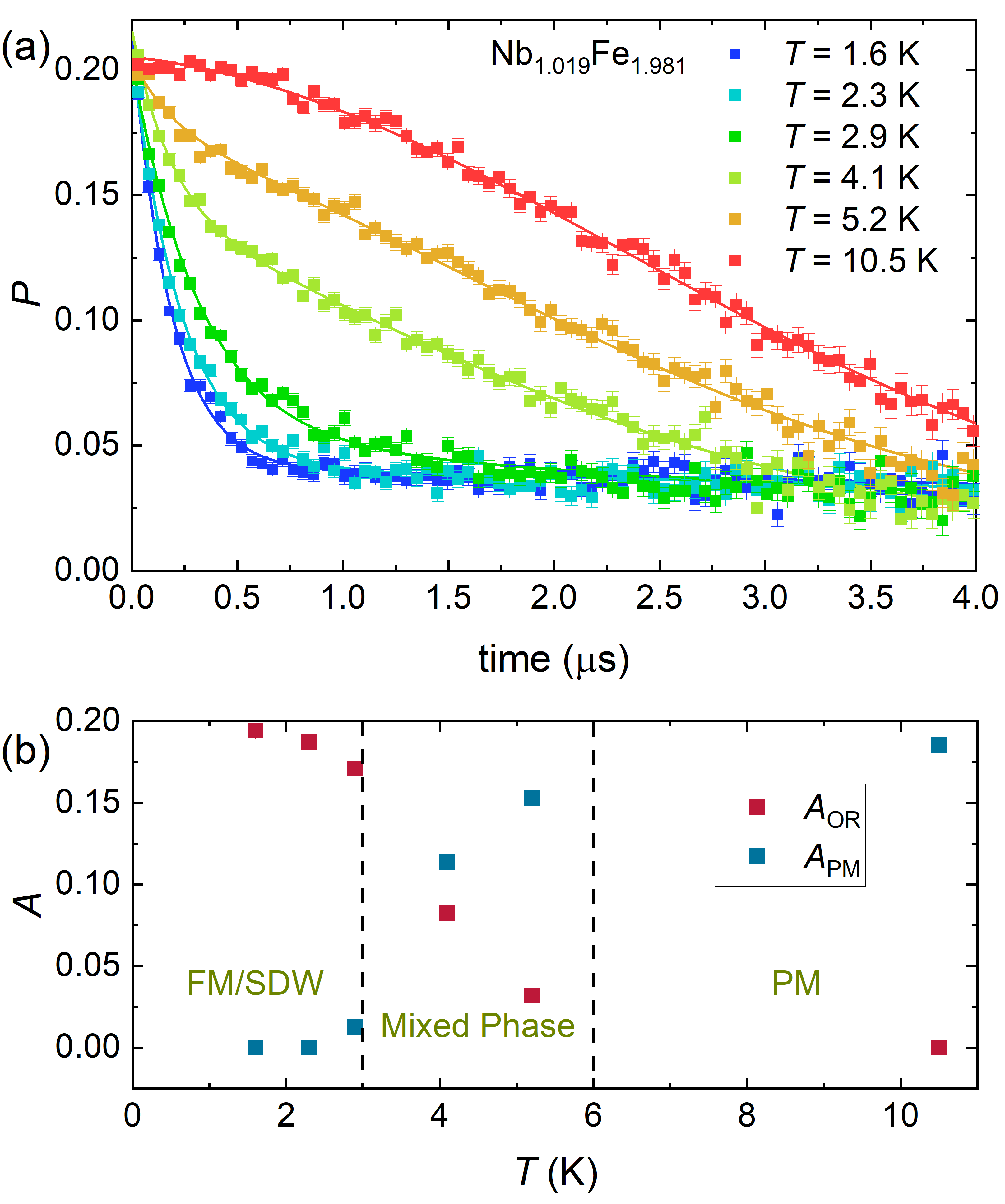}
		\caption{\label{C2} (a) $\mu$SR time spectra of a ZF measurement on Nb$_{1.019}$Fe$_{1.981}$ for temperatures between $1.6$ and $10.5$\,K and (b) asymmetries $A_{\mathrm{PM}}$ and $A_{\mathrm{OR}}$ as function of temperature derived from these measurements; for details, see text.}
	\end{figure}
	
	To complete the analysis of Nb$_{1-y}$Fe$_{2+y}$ using $\mu$SR measurements, we carried out ZF experiments on two polycrystalline samples with compositions close to the quantum critical regime: Nb$_{1.019}$Fe$_{1.981}$ and Nb$_{1.0055}$Fe$_{1.9945}$, and start to discuss the measurements on first composition. Previously, a ferromagnetic phase was proposed to exist for the Nb-rich samples at low temperatures \cite{Moroni-Klementowicz2009}. To investigate this proposal, in Fig. \ref{C2} (a) we depict the $\mu$SR time spectra of a ZF measurement on Nb$_{1.019}$Fe$_{1.981}$ for temperatures between $1.6$ and $10.5$\,K. The asymmetry functions for low temperatures are strongly damped, as can be seen by the steep drop of the polarization signal at short times. Most likely, this is caused by long-range magnetic order. For higher temperatures of $4.1$ and $5.2$\,K at approximately $0.5$\,$\mathrm{\mu s}$ a kink-like point is visible, where the signal starts to level off. In contrast, at $10.5$\,K the damping of the asymmetry signal is significantly weaker. 
	
	To study the magnetic behavior of Nb$_{1.019}$Fe$_{1.981}$ in detail, we fit the experimental data using
	\begin{equation}
		\label{fit_C2}
		P = A_{\mathrm{PM}} F_{\mathrm{GKT}} e^{-\lambda t} + A_{\mathrm{OR}} F_{\mathrm{GKT}} F_{\mathrm{FM}} + A_{\mathrm{BG}} ,
	\end{equation}
	(solid lines in Fig. \ref{C2} (a)), with the physical models associated to these functions introduced in Sec. \ref{sec:level2}. Since no oscillations are detected even at temperatures of $1.6$\,K, the frequency $\nu$ in $F_{\mathrm{FM}}$ was set to zero. Since the magnetic moments are so small that the frequency can be set to zero, it is reasonable to assume only one muon stopping site in the unit cell. This again allows the free parameters in the fit to be reduced.

	The temperature dependence of the asymmetries $A_{\mathrm{PM}}$ and $A_{\mathrm{OR}}$ (Fig. \ref{C2} (b)) can be used to determine the magnetic state of the sample. Starting at high temperatures $\sim 10$\,K, the best fit result is achieved with $A_{\mathrm{OR}} = 0$, so that the $\mu$SR time spectra is fully described by $F_{\mathrm{GKT}}$. The sample is thus in the paramagnetic phase and the $\mu$SR signal is only controlled by the randomly oriented nuclear magnetic moments. With decreasing temperature $A_{\mathrm{PM}}$ slowly decreases and $A_{\mathrm{OR}}$ increases. The sample seems to be in a partially ordered mixed phase, with $A_{\mathrm{PM}}$ and $A_{\mathrm{OR}}$ of similar size at around 4\,K. In addition, in this phase the exponential damping rate $\lambda$ increases with decreasing temperature from $0.05(1)$\,$\mathrm{\mu s^{-1}}$ at $10.5$\,K to $0.33(5)$\,$\mathrm{\mu s^{-1}}$ at $2.9$\,K. This indicates the appearance of magnetic fluctuations close to the magnetic transition. The temperature dependence of $\lambda$ is similar in this sample (above $\sim 3$\,K) and in the quantum critical sample Nb$_{1.0117}$Fe$_{1.9883}$ (above $\sim 0.7$\,K). Despite the sparse data density for Nb$_{1.019}$Fe$_{1.981}$, here $\lambda$ also appears to follow Eq. (\ref{fit_lambda}) with $\alpha \sim 1$.
	
	Upon further lowering the temperature the signatures of long-range magnetic order dominate the $\mu$SR spectra and $A_{\mathrm{PM}}$ approaches 0. Globally, this result is in good agreement with the published phase diagram (see Fig. \ref{PD}). However, our data indicate that there is no sharp phase boundary, but a gradual transition from the magnetically ordered to the paramagnetic phase as function of temperature.
	
	\begin{figure}[t]
		\includegraphics[scale=0.148]{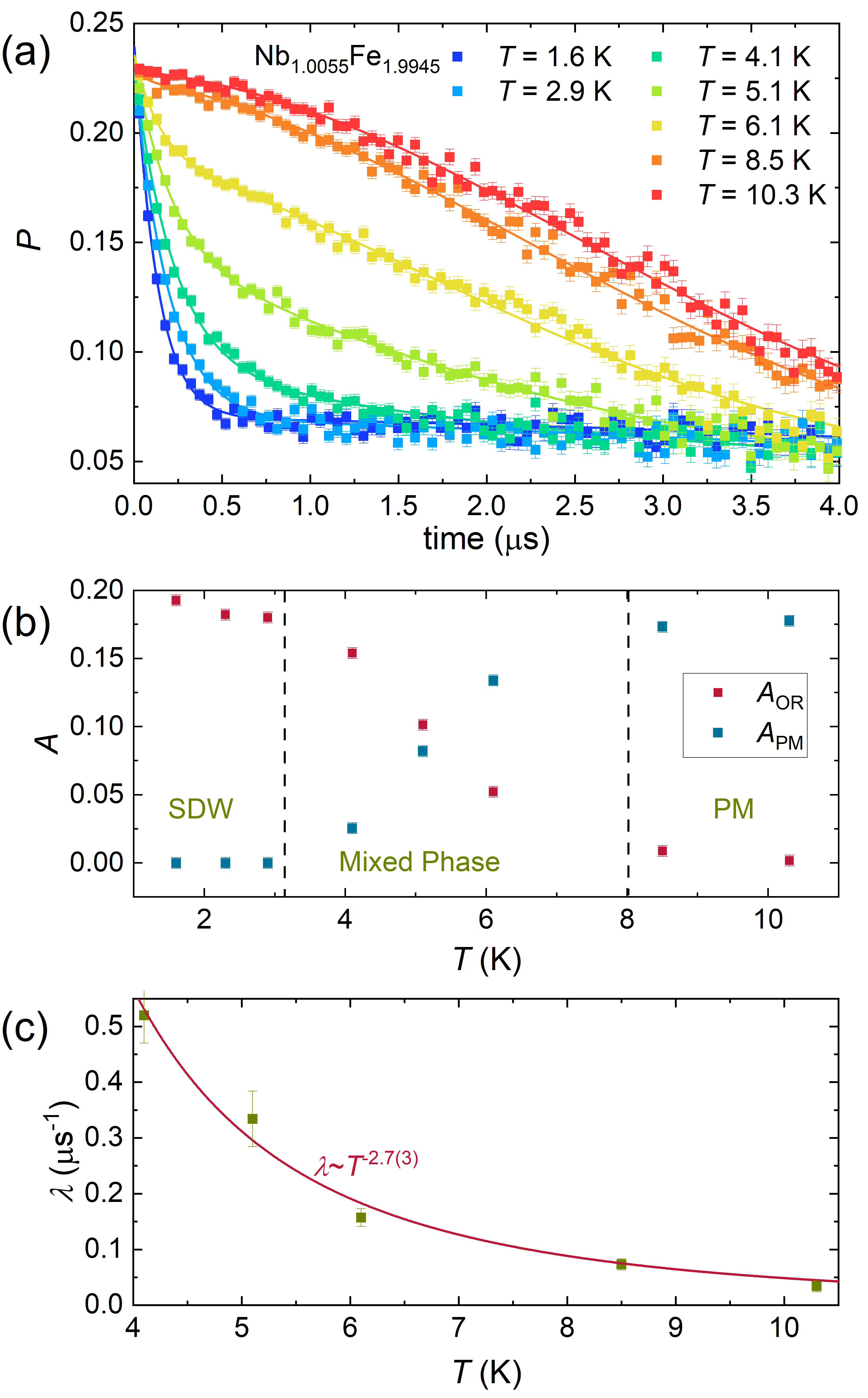}
		\caption{\label{C4} (a) $\mu$SR time spectra of a ZF measurement on Nb$_{1.0055}$Fe$_{1.9945}$ for temperatures between $1.6$ and $10.3$\,K, (b) the temperature dependencies of the asymmetries $A_{\mathrm{PM}}$ and $A_{\mathrm{OR}}$ and (c) the exponential depolarization rate $\lambda$; for details, see text.}
	\end{figure}
	
	The question of the type of magnetic order of Nb$_{1.019}$Fe$_{1.981}$ remains to be addressed. At this point, there is no firm verification from a microscopic experimental technique for the claimed ferromagnetic nature of the ordered phase for compositions with $y < -0.015$. As well, it could be possible that the phase FM1 (Fig. \ref{PD}) contains an additional SDW phase near the quantum critical point similar to the Fe-rich regime of the phase diagram. In principle, for the proposal of a ferromagnetic phase an oscillating behavior would be expected in our $\mu$SR spectra. However, in Nb$_{1.019}$Fe$_{1.981}$ the magnetic moments likely are so small that the signal just drops steeply and no oscillations are visible. It makes a definite statement about the type of magnetic order difficult. In fitting our data, we also tried to adjust the fit approach by using instead of function $F_{\mathrm{FM}}$ [Eq. (\ref{ferro})] for ferromagnetic signals the function $F_{\mathrm{SDW}}$ [Eq. (\ref{SDW})] for a SDW signal at $1.6$\,K. Ultimately, these two approaches do not produce a significant quality difference of the fits. Therefore, it is not clear whether Nb$_{1.019}$Fe$_{1.981}$ is in a FM or SDW phase at low temperatures. Another approach is to compare the spectra shown here with measurements on an Fe-rich sample where the SDW phase is well established.
	
	Therefore, we present the ZF $\mu$SR measurements on Nb$_{1.0055}$Fe$_{1.9945}$, {\it i.e.}, at the Fe-rich side of the phase diagram next to quantum criticality. For a sample with similar composition a SDW ground state was discovered previously \cite{Rauch2015} and a transition temperature of approximately $5$\,K follows from the phase diagram \cite{Niklowitz2019}. In Fig. \ref{C4}(a) we present $\mu$SR measurements between $1.6$\,K and $10.3$\,K for Nb$_{1.0055}$Fe$_{1.9945}$. In general, the behavior of the $\mu$SR signal is very similar to that of Nb-rich Nb$_{1.019}$Fe$_{1.981}$ discussed above. At low temperatures and short times a steep drop of the polarization signal can be seen due to the long-range magnetic order. At higher temperatures the initial steep drop becomes less pronounced until it is not visible any more at $8.5$\,K.
	
	To analyze the measurements, the same fit approach as in Eq. (\ref{C2}) for Nb$_{1.019}$Fe$_{1.981}$ is used. Since it is known that this sample is in a SDW phase at low temperatures, the function $F_{\mathrm{SDW}}$ [Eq. (\ref{SDW})] was chosen instead of $F_{\mathrm{FM}}$. At high temperatures the $\mu$SR signal can again be fully described by the first term with $A_{\mathrm{PM}}$ (Fig. \ref{C4}(b)). Accordingly, the sample is not yet magnetically ordered and resides in the paramagnetic phase. With lowering the temperature $A_{\mathrm{PM}}$ starts to decrease and $A_{\mathrm{OR}}$ to increase. It implies that phase separation sets in and parts of the sample begin to order magnetically. The intersection of $A_{\mathrm{PM}}$ and $A_{\mathrm{OR}}$ is at $5$\,K and fits very well the boundary of the phase diagram. Below $3$\,K the damping of the $\mu$SR signal can entirely be fitted with the second term of Eq. (\ref{fit_C2}) and thus, the sample is now fully ordered in the SDW phase.
	
	Since the description of the $\mu$SR signal improves by adding the exponential damping term, it can be assumed that electronic spin fluctuations play an important role. The temperature dependence of the exponential depolarization rate $\lambda$ (Fig. \ref{C4}(c)) is therefore of particular interest. At high temperatures $\lambda$ is close to zero and increases with decreasing temperature. As for the other samples close to or at quantum criticality, this is quantified with a simple power-law fit: $\lambda \sim T^{-2.7(3)}$. The increase in $\lambda$ with decreasing temperature can be interpreted as arising from a slowing down of the spin fluctuations. 
	
	This analysis demonstrates that Nb$_{1.019}$Fe$_{1.981}$ and Nb$_{1.0055}$Fe$_{1.9945}$ qualitatively and semi-quantitatively behave quite similarly. Since the SDW phase is well established in Nb$_{1.0055}$Fe$_{1.9945}$, it remains a possibility that there is also a SDW phase in the Nb-rich regime.
	
	\section{Discussion}
	
	\begin{figure}[t]
		\includegraphics[scale=0.34]{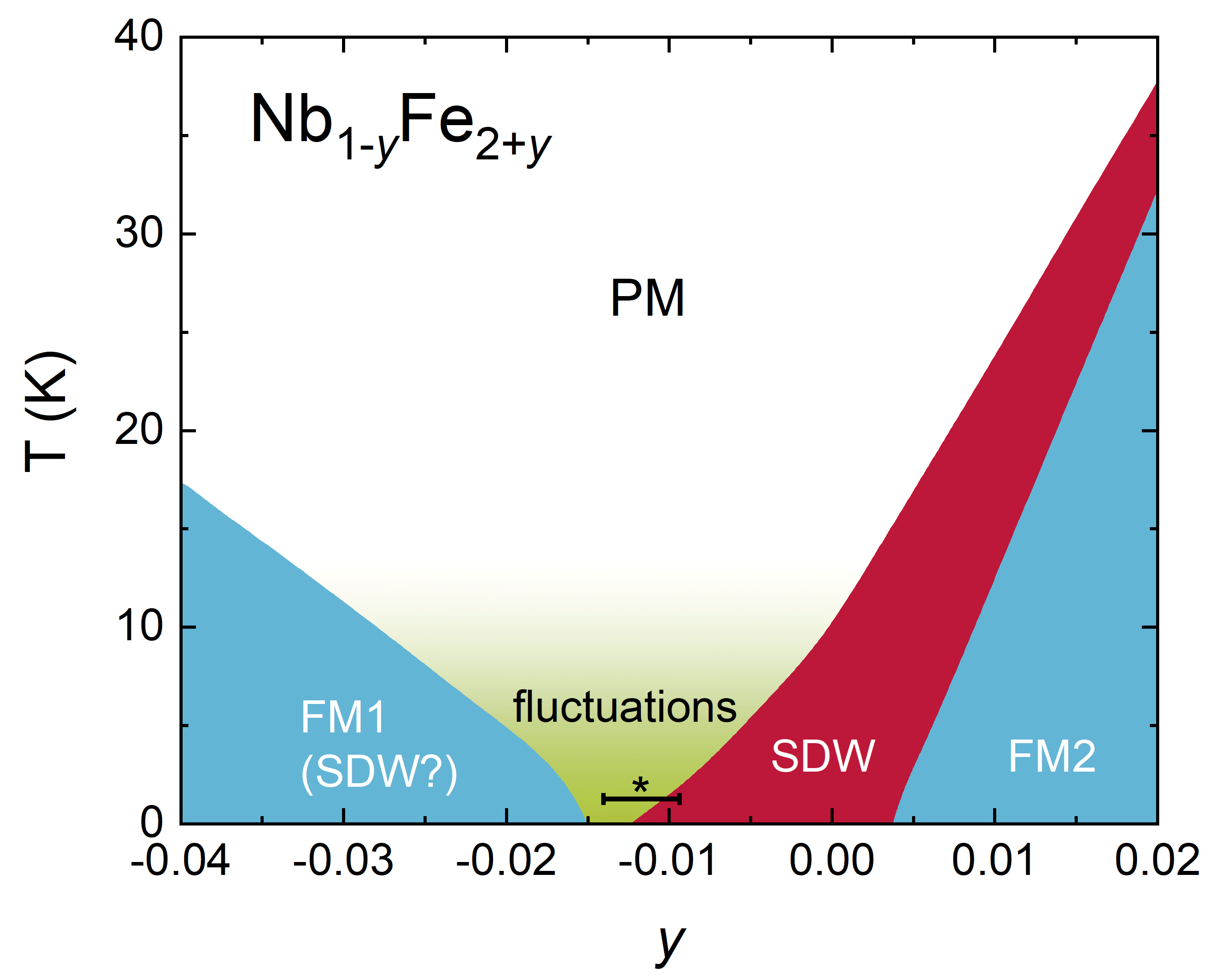}
		\caption{\label{PD_neu} Phase diagram of Nb$_{1-y}$Fe$_{2+y}$ with the ferromagnetic ultralow-moment phases (FM1, FM2) and the spin density phase (SDW) after Niklowitz \textit{et al.} \cite{Niklowitz2019}. In addition, we have marked the region of prominent (quantum critical) magnetic fluctuations as the green shaded area. The bar marked with a star shows the statistical variation of the local stoichiometry of Nb$_{1.0117}$Fe$_{1.9883}$; for details, see text.}
	\end{figure}

	Summarizing our findings, we have proven the existence of two distinct muon stopping sites in the ferromagnetically ordered phase of Nb$_{1-y}$Fe$_{2+y}$. Our experiments and simulations show that the muon stopping site is within the tetrahedra consisting of three Fe(2a) and one Nb(4f) atoms. A detailed analysis of the ZF $\mu$SR measurements on Nb$_{0.982}$Fe$_{2.018}$ confirms the existence of the SDW phase, as well as the low-moment FM phase in the Fe-rich regime of the phase diagram.
	
	Of special interest, and to this point only superficially characterized by microscopic techniques, is the part of the phase diagram where no long-range magnetic order has been detected. We have analyzed ZF and LF $\mu$SR measurements on quantum critical Nb$_{1.0117}$Fe$_{1.9883}$, and have studied long-range ordered samples close to the quantum critical regime, \textit{i.e.}, Nb$_{1.0055}$Fe$_{1.9945}$ and Nb$_{1.019}$Fe$_{1.981}$. Taken together with the data from Ref. \cite{Rauch2015}, we have thus fully characterized the magnetic phase diagram of Nb$_{1-y}$Fe$_{2+y}$ in the weakly magnetic/quantum critical range by the local probe muon spin rotation and relaxation $\mu$SR.
	
	No long-range magnetic order was detected down to $19$\,mK for Nb$_{1.0117}$Fe$_{1.9883}$ in our measurements. Instead, the $\mu$SR time spectra are dominated by the influence of static nuclear dipoles above $10$\,K. Upon cooling, the damping of the $\mu$SR signal significantly increases due to the slowing down of dynamic magnetic fluctuations from the electron system. Magnetic fluctuations similar in character were also detected for Nb$_{1.0055}$Fe$_{1.9945}$ and Nb$_{1.019}$Fe$_{1.981}$ just above their transition temperatures. Also, the temperature dependence of the exponential depolarization rate $\lambda$ is similar for all three compositions. Taking these observations together, we can identify a region of prominent magnetic fluctuations with the relaxation rate $\lambda \propto T^{-\alpha}$ in the phase diagram of Nb$_{1-y}$Fe$_{2+y}$, as seen by the technique of $\mu$SR (green shaded range in Fig. \ref{PD_neu}). Moreover, while we can not provide a definite proof, our data would be consistent with the existence of a SDW phase also on the Nb-rich long-range ordered region of the phase diagram.
	
	Regarding our observations in the alloying range of Nb$_{1-y}$Fe$_{2+y}$ with quantum critical fluctuations, we will compare our data to those from $\mu$SR studies on other materials at quantum criticality. Prominent examples are CeFePO \cite{Lausberg2012}, UCu$_4$Pd \cite{MacLaughlin1998,MacLaughlin2000}, CePd$_{1-x}$Rh$_x$ \cite{Adroja2008} and YFe$_2$Al$_{10}$ \cite{Huang2018}. For the first three materials, in contrast to our case, the low temperature data have been fitted using stretched exponentials. This approach has been introduced in the context of spin glass studies \cite{Campbell1994a,Campbell1994b}, where it is associated to the spin autocorrelation function. In spin glasses the physical picture is based upon the existence of a wide spectrum of local spin fluctuation rates. While it may be assumed that also in quantum critical systems a wide spectrum of spin fluctuation rates exists, it is not clear if the same physics truly applies in inherently disordered spin glasses as in dense magnetic systems. Therefore the fit procedure is considered to be a phenomenological approach. Moreover, it is known that stretched exponentials might misrepresent or mask underlying behavior such as the existence of multiple, well-defined relaxation time scales \cite{Heffner2000}. In certain cases stretched exponentials have explicitly been used to facilitate an effective description of a wide distribution of relaxation rates in disordered quantum critical materials \cite{MacLaughlin2001}. 
	
	After all, the distinctly different behavior of Nb$_{1-y}$Fe$_{2+y}$ at quantum criticality sets our material apart from these cases. In particular, the exponential muon relaxation in Nb$_{1-y}$Fe$_{2+y}$ suggests that structural disorder is not an issue to the same degree as for UCu$_4$Pd, CePd$_{1-x}$Rh$_x$ or even CeFePO (where a spin glass transition has been detected in single crystalline material) \cite{MacLaughlin1998,Adroja2008,Lausberg2012}, and that we deal with a comparatively clean magnetic system. Conceptually, this observation might simply reflect that in itinerant magnets - because of the spatially extended electronic states - crystallographic disorder is less an issue than in local moment systems, as for instance demonstrated by the insensitivity of skyrmionic behavior to crystallographic disorder in Co$_8$Zn$_8$Mn$_{4-x}$Fe$_x$ \cite{menzel2019}.
	
	Instead, it is more instructive to compare our results to those reported for YFe$_2$Al$_{10}$, a material believed to reside close to a ferromagnetic quantum critical point \cite{Huang2018}. Broadly speaking, our experimental observations at quantum criticality for Nb$_{1.0117}$Fe$_{1.9883}$ exhibit similarities to the findings for YFe$_2$Al$_{10}$. For both systems, at high temperatures a simple Gauss-Kubo-Toyabe function with some exponential damping describes the experimental data. It corresponds to a paramagnetic system, where muon precession arises from quasistatic nuclear dipolar fields, with some electronic relaxation. As temperature is lowered, for both systems the exponential damping becomes more pronounced, with $\lambda$ rising by an order of magnitude, but flattening off below a crossover temperature.
	
	Quantitatively, these observations reveal a difference of energy scales between Nb$_{1.0117}$Fe$_{1.9883}$ and YFe$_2$Al$_{10}$ by an order of magnitude. While in the former system the GKT character is prominent down to $\sim$ 10\,K, with the relaxation then taking over with lowering temperature, in YFe$_2$Al$_{10}$ this occurs closer to 1\,K. In Nb$_{1.0117}$Fe$_{1.9883}$ the significant change by more than an order of magnitude of the relaxation rate occurs in the temperature window $\sim$ 10 to 1\,K, in contrast in YFe$_2$Al$_{10}$ between $\sim$ 1 and 0.1\,K. As well, the crossover temperatures to a behavior $\lambda \propto T^{-\alpha}$ with a small value $\alpha$ differ, with $\sim$ 0.7\,K in Nb$_{1.0117}$Fe$_{1.9883}$ compared to 0.1\,K in YFe$_2$Al$_{10}$. Aside from these quantitative differences, the temperature dependence of $\lambda$ itself is very similar for both materials, with at higher temperatures $\lambda \propto T^{-1.16}$ in Nb$_{1.0117}$Fe$_{1.9883}$ quite close to $\lambda \propto T^{-1.2}$ in YFe$_2$Al$_{10}$. As well, below the crossover in YFe$_2$Al$_{10}$ $\lambda$ saturates, while in our case we have a residual finite, but very small value $\alpha = 0.15(2)$.
	
	In detail, there is one particular difference between the two systems: While in YFe$_2$Al$_{10}$ muon depolarization was parameterized entirely by using a GKT-function and exponential relaxation, for Nb$_{1.0117}$Fe$_{1.9883}$ it was necessary to introduce a second fit component with an exponential Kubo-Toyabe term. As pointed out, the necessity to introduce this term is directly seen in the experimental data and suggests that in part of the sample magnetic fluctuations from dilute electronic dipolar moments are governing muon relaxation. The absence of such a term in the muon data of YFe$_2$Al$_{10}$ could reflect some qualitative difference in the fluctuation spectra in the two systems. We note however, given the apparent difference in energy scales between Nb$_{1.0117}$Fe$_{1.9883}$ and YFe$_2$Al$_{10}$, that it is conceivable that analogous behavior to Nb$_{1.0117}$Fe$_{1.9883}$ would appear in YFe$_2$Al$_{10}$ in a more narrow and much harder to study temperature range between 0.1 and 1\,K, and might have not been detected for reasons of experimental resolution.
	
	With this scientific context for our study on Nb$_{1-y}$Fe$_{2+y}$, we will now construct an ''experimental reading'' of how quantum critical behavior appears in the microscopic and local probe $\mu$SR. First, our parametrization of the data on Nb$_{1.0117}$Fe$_{1.9883}$ suggests a contribution that corresponds to 30\,\% of the sample volume of muon relaxation from magnetic fluctuations of dilute dipolar moments. A possible origin of such a contribution might be structural inhomogeneity in our samples Nb$_{1-y}$Fe$_{2+y}$. With respect to sample quality, it has been demonstrated for Nb$_{1-y}$Fe$_{2+y}$ that stoichiometric variations in a given sample are negligible \cite{Brando2008,Moroni-Klementowicz2009,Pfleiderer2010,Rauch2015}. It implies that in the alloying phase diagram of Nb$_{1-y}$Fe$_{2+y}$, at least for small values $y$, there is no miscibility gap causing local stoichiometry variations \cite{Moroni-Klementowicz2009,Suellow1994}. In single crystals, it has been observed that a sudden change of local magnetic properties might occur along the growth axis, as a result of preferential evaporation of one element during the growth process leading to a stoichiometry gradient \cite{Pfleiderer2010}. Our data, in contrast, indicate that the second phase volume seen by the muons emerges as function of temperature out and on behalf of the first phase volume, {\it i.e.}, it has the character of a spontaneous phase separation in an otherwise homogeneous material. Therefore, our data are inconsistent with the notion of a distinct and separate phase volume in our quantum critical sample Nb$_{1.0117}$Fe$_{1.9883}$.

	Alternatively, a natural source for a distribution of local magnetic environments in Nb$_{1-y}$Fe$_{2+y}$ might be attributed to inherent local statistical stoichiometry variations, leading to a distribution of local compositions in the real material. At present, there is no direct information about the type or level of site exchange in Nb$_{1-y}$Fe$_{2+y}$. Still, we can consider a simple alloying model by assuming that stoichiometric NbFe$_2$ is crystallographically perfectly ordered. Then, for Nb$_{1-y}$Fe$_{2+y}$, $y \neq 0$, all off-stoichiometric atoms are built into the crystallographic lattice through Nb/Fe site exchange. In the Laves phase structure ({\it i.e.}, the unit cell is formed by four chemical formula units AB$_2$) each off-stoichiometric Nb atom defines a crystallographic cluster around its position of $N$ unit cells, with $N = 25 \times 100 |y| $, in which no second off-stoichiometric Nb atom will be found on average. As a statistical distribution, there is some uncertainty to the value of $N$, {\it i.e.}, $\Delta N = \sqrt{N}$, resulting in the typical cluster size for a given value $y$ of $N \pm \Delta N$ unit cells around each off-stoichiometric Nb atom. 
	
	Applying this to quantum critical Nb$_{1.0117}$Fe$_{1.9883}$, we find a cluster size (in terms of the number of unit cells) $N \pm \Delta N = 29 \pm 5$. Such a distribution of cluster sizes ranging from 24 to 34 implicates a statistical variation of the local stoichiometry $y$, {\it i.e.}, it varies locally between between $-0.014 \leq y \leq -0.0096$. In Fig. \ref{PD_neu}, we have indicated this stoichiometry range around the nominal composition of the quantum critical sample, revealing that the local composition of this sample is expected to stretch into the SDW regime on the Fe rich side of the phase diagram. 
	
	It is far from clear how such a statistically induced stoichiometry distribution translates into the magnetic behavior of a material. As pointed before, NbFe$_2$ is an itinerant magnet, where local effects/inhomogeneities should have less an influence on the magnetic properties. On the other hand, the very strong dependence of the magnetic phase diagram on the level of alloying implies that the local composition needs to be taken into account. It is likely that those parts of Nb$_{1.0117}$Fe$_{1.9883}$ with a local stoichiometry from the SDW range of the phase diagram - even if not resulting in long-range magnetic order because of size effects/limited correlation length - exhibit a different dynamic behavior than those from the quantum critical range.
	
	Therefore, we propose a phenomenological picture of quantum criticality in Nb$_{1.0117}$Fe$_{1.9883}$ which is based upon a very narrow, but finite range of local compositions close to the nominal one - conceptually, compared to the materials discussed before this corresponds to a ''weak residual disorder picture''. Within this view, the GKT behavior at 10\,K and above represents - on the time scale of the muon experiment $\sim \mu s$ - the ordinary paramagnetic regime. Quantum critical fluctuations, revealed by an enhanced simple exponential muon depolarization rate $\lambda$, take over in the temperature range $\sim 1$ to 10\,K. Experimentally, it appears that for comparatively clean systems such as Nb$_{1.0117}$Fe$_{1.9883}$ or YFe$_2$Al$_{10}$ the depolarization rate essentially evolves with $\lambda \propto T^{-\alpha}$ with $\alpha \sim 1$. In Nb$_{1.0117}$Fe$_{1.9883}$, as temperature is lowered these quantum critical fluctuations compete with fluctuations from dilute dipolar moments. In our reading these would stem from magnetic clusters from the SDW compositional range of the magnetic phase diagram, and which start to become relevant below the ordering temperature of the SDW phase (a few K). Thus, at low temperatures $\sim 1$\,K there is a coexistence of quantum critical fluctuations in the bulk with those from larger magnetic clusters, with the first accounting for 70 \% of the sample volume and the latter for 30 \%. Experimentally, we observe that the volume fraction from dilute dipolar moments becomes constant below $\sim 1$\,K, consistent with the view of a narrow compositional distribution, with only part of it in the SDW compositional range. Finally, we find a crossover at $\sim 0.7$\,K in the depolarization rate $\lambda \propto T^{-\alpha}$ towards a much smaller value $\alpha \sim 0.1$. It is unclear if there is a relationship to the dilute dipolar volume becoming constant, or if it is inherent to the fluctuation spectrum. In favor of the latter view is the observation that a similar behavior is seen for YFe$_2$Al$_{10}$, but clearly our understanding here is very limited. 
	
	The interpretation of the $\mu$SR data for Nb$_{1.0117}$Fe$_{1.9883}$ appears to be consistent with the concept of a statistical variation of the local stoichiometry. Still, it might be hypothesized that there is a more exotic explanation for the occurrence of the 30\,\% volume fraction in the $\mu$SR signal. It could be argued that the spin fluctuation spectrum at quantum criticality possesses a very broad spectral character, and that part of the spectrum appears static on the time scale of $\mu$SR. This view, however, is not easily reconciled with the necessity to choose two distinct magnetic volumes at low temperatures in our fit. 
		
	There are two more takeaways from our study. Conceptually, the phenomenological picture we have developed borrows from the Griffiths phase scenario. Originally, in UCu$_4$Pd, one (heuristical) approach to explain the $\mu SR$ and NMR experiments was based on the appearance of a Griffiths phase \cite{MacLaughlin1998}. In the Griffiths phase scenario magnetic impurities lead to magnetic fluctuations on a range of length and energy scales. It results in the formation of a magnetic phase consisting of disordered spin clusters of different size \cite{Griffiths1969,Neto1998}, that gradually fill the sample volume as temperature is lowered. The magnetic dynamical behavior of the spin clusters leads to a susceptibility that diverges at low temperatures with $\chi(T)\propto T^{-1+\eta}$, where $\eta<1$ \cite{Neto1998,Neto2000,lameta}. On approaching a quantum critical point, calculations from Ref. \cite{Neto2005} suggest that $\eta \rightarrow 0$ and thus $\chi(T)\rightarrow T^{-1}$.  
	
	Within the Griffiths phase scenario the full volume of a quantum critical sample develops such spin cluster fluctuations, while in our case we experimentally determine that at best 30 \% of the sample volume might contribute - in other words, our sample Nb$_{1.0117}$Fe$_{1.9883}$ would be in a ''weak Griffiths phase disorder limit''. Since the damping of the $\mu$SR spectra is considered a measure of the magnetic susceptibility, we compare the predicted behavior of the susceptibility with the temperature dependence of $\lambda$ from the quantum critical sample Nb$_{1.0117}$Fe$_{1.9883}$: As demonstrated, $\lambda$ follows a simple power law $\lambda\propto T^{-\alpha}$ between $0.7$ and $10$\,K with $\alpha=1.16(9)$, reminiscent of the susceptibility predicted by theory for a clean sample at quantum criticality. This observation raises the question if there is still some Griffiths phase physics associated to the quantum critical point in Nb$_{1.0117}$Fe$_{1.9883}$ and which will require further studies.
	
	Secondly, the difference in behavior between, on the one hand, systems like UCu$_4$Pd, CePd$_{1-x}$Rh$_x$ and CeFePO and, on the other hand, Nb$_{1.0117}$Fe$_{1.9883}$ and YFe$_2$Al$_{10}$ lends itself to formulate a simple classification scheme: While for the first materials the quantum critical behavior should be considered to occur in a strongly disordered environment, for the latter systems disorder plays a limited or possibly no role, {\it viz.}, clean quantum criticality.
	
	\section{Conclusion} 
	
	Altogether, from our $\mu$SR study, we have thoroughly characterized the full magnetic phase diagram of Nb$_{1-y}$Fe$_{2+y}$. In a first step, the muon stopping site in the crystallographic lattice of NbFe$_2$ was established. The central focus of this paper was the investigation of a sample from the quantum critical regime using the microscopic technique $\mu$SR. We have demonstrated that magnetism at lowest temperatures is dominated by magnetic fluctuations of different character. Aside from quantum critical fluctuations, part of these might be related to natural stoichiometry variations causing fluctuations from dilute dipolar moments. We speculate that these might show characteristics of a Griffiths phase. As Nb$_{1-y}$Fe$_{2+y}$ is a comparatively clean magnetic system, our observations likely reflect the intrinsic behaviour of clean materials close to a quantum critical point. Moreover, from a more general perspective, the technical approach we have chosen to resolve the microscopic details of the magnetic phase diagram of Nb$_{1-y}$Fe$_{2+y}$ may serve as a blueprint for future work on related topics.

	\begin{acknowledgments}
	We thank D. Moroni and W.J. Duncan for the preparation of many of the samples used in this work \cite{Moroni-Klementowicz2009,Duncan2010} and A. Neubauer and C. Pfleiderer for their help with preparing the single crystal Nb$_{0.982}$Fe$_{2.018}$ \cite{Friedemann2013,Duncan2010}.	This work is based on experiments performed at the Swiss Muon Source S$\mu$S, Paul Scherrer Institute, Villigen, Switzerland. E.S. and F.J.L. acknowledge support from DFG project Li 244/12-2. E.S. also acknowledges support by DFG through the projects C06 and C09 of the SFB 1143 (project-id 247310070) and the W\"urzburg-Dresden Cluster of Excellence on Complexity and Topology in Quantum Matter–ct.qmat (EXC 2147, project-id 390858490). The work was supported by the EPSRC of the UK (Grants No. EP/K012894/1 and No. EP/ P023290/1).
	\end{acknowledgments}


\begin{thebibliography}{00}
\bibitem{lameta} In the literature, $\lambda$ is often used instead of $\eta$. Here, we use $\eta$ to avoid confusion with the depolarization rate.
\end{thebibliography}
\end{document}